\documentstyle[11pt]{article}

\def\proj{E\hskip -.69em I}     
\begin{document}
\begin{titlepage}
\begin{centering}
 
{\ }\vspace{1cm}
 
{\Large\bf Quantisation without Gauge Fixing:}\\
\vspace{0.5cm}
{\Large\bf Avoiding Gribov Ambiguities through the Physical Projector}\\
\vspace{2.5cm}
Victor M. Villanueva$^{\ddag ,}$\footnote{Work done while at the
{\em Instituto de F\'{\i}sica, Universidad de Guanajuato, 
37150 Le\'on, M\'exico}}$^{,}$\footnote{E-mail: 
{\tt vvillanu@ifm1.ifm.umich.mx}},
Jan Govaerts$^{\dag ,}$\footnote{E-mail: {\tt govaerts@fynu.ucl.ac.be}}
and Jose-Luis Lucio-Martinez$^{\star ,}$\footnote{E-mail: 
{\tt lucio@ifug3.ugto.mx}}\\
\vspace{0.6cm}
$^{\ddag}${\em Instituto de F\'{\i}sica y Matem\'aticas}\\
{\em Universidad Michoacana de San Nicol\'as de Hidalgo}\\
{\em P.O. Box 2-82, Morelia Michoac\'an, M\'exico}\\
\vspace{0.5cm}
$^{\dag}${\em Institut de Physique Nucl\'eaire, 
Universit\'e catholique de Louvain}\\
{\em 2, Chemin du Cyclotron, B-1348 Louvain-la-Neuve, Belgium}\\
\vspace{0.5cm}
$^{\star}${\em Instituto de F\'{\i}sica, Universidad de Guanajuato}\\
{\em P.O. Box E-143, 37150 Le\'on, M\'exico}
\vspace{2cm}
\begin{abstract}

\noindent The quantisation of gauge invariant systems usually proceeds through
some gauge fixing procedure of one type or another. Typically for most cases,
such gauge fixings are plagued by Gribov ambiguities, while it is only for
an admissible gauge fixing that the correct dynamical description of the
system is represented, especially with regards to non perturbative phenomena. 
However, any gauge fixing procedure whatsoever may
be avoided altogether, by using rather a recently proposed new approach
based on the projection operator onto physical gauge invariant states only,
which is necessarily free on any such issues. These different aspects of
gauge invariant systems are explicitely analysed within a solvable
U(1) gauge invariant quantum mechanical model related to the dimensional
reduction of Yang-Mills theory.

\end{abstract}

\vspace{10pt}

\end{centering} 

\vspace{30pt}


\vspace{25pt}

\noindent September 1999

\end{titlepage}

\setcounter{footnote}{0}

\section{Introduction}
\label{Sect1}

As beautiful, elegant and powerful as is the general principle of 
local gauge invariance, there is also a price to be paid,
especially when it comes to quantising such theories. Indeed, the
characteristic feature of systems possessing such symmetries is the
presence among their degrees of freedom of redundant gauge variant
variables required for a manifest realisation of the gauge invariance
principle, and possibly also of other symmetries such as spacetime covariance.
Consequently, the actual physical configuration space of these systems
is typically quite intricate, whose non trivial topology is at the
heart of fundamental non perturbative phenomena responsible for the
rich physics implied by these theories.

The actual physical configuration space, being parametrised by the
initial degrees of freedom modded out by the set of local (large and small)
gauge transformations, corresponds to the set of gauge orbits in the
configuration space of initial degrees of freedom.
Usually the quantisation of such systems implements first some
gauge fixing procedure of one kind or another, which should in principle select 
among the initial degrees of freedom a single representative of each one
of all the gauge orbits accessible to the system throughout its dynamical 
time evolution, thereby defining an admissible gauge fixing procedure. 
However, most gauge fixing procedures do not meet that requirement, and
are then said to be plagued by a Gribov ambiguity\cite{Gribov,Singer}.
In fact, two types of Gribov ambiguities ought to be 
distinguised\cite{Gov1}. The first type of Gribov problem, 
or ``local Gribov problem", arises when a gauge
fixing procedure selects more than one representative of the same
gauge orbit, the important point being however that
the total number of these representatives must be summed by also accounting
for the oriented integration measure induced on the physical configuration 
space of gauge orbits by the local integration measure on the initial
configuration space\cite{Gov1,Hirschfeld}. 
The second type of possible Gribov problem, or ``global Gribov problem", arises
when not all gauge orbits of the system are selected through some
gauge fixing procedure\cite{Singer}. Clearly, an admissible gauge fixing 
procedure is one which does not suffer neither local nor global Gribov problems.

In spite of the importance of the issue, especially when it comes to 
non perturbative phenomena, the global Gribov problem is usually not 
specifically considered in the literature, though it is a possibility which
often arises\cite{Singer,Gov1}. A local Gribov problem on the other
hand, arises typically when the so-called Faddeev reduced phase space 
approach\cite{Gov1} for example is developed towards the quantisation of a gauge
invariant system\footnote{Note that a {\em non vanishing\/} Faddeev-Popov
determinant establishes the absence of a local Gribov problem only
but not necessarily of a global one, and then 
{\em only for infinitesimal gauge transformations\/} but not necessarily for
finite (small and even large) ones. Thus, a non vanishing Faddeev-Popov
determinant does not necessarily define an admissible gauge fixing
procedure in the sense considered above\cite{Gov1}, contrary to the common
usage of the word in the literature.}, even in the simplest cases\cite{Gov1}.
The original example of a Gribov problem\cite{Gribov} could be of this type, 
but when accounting for the oriented integration measure over the space of 
gauge orbits, it may well be that the original Gribov example of gauge 
redundancy in a gauge fixing procedure is not a Gribov problem after 
all\cite{Hirschfeld}.
In fact, Gribov's suggestion for a resolution of local Gribov problems,
by res\-tric\-tion to the fundamental domain of non vanishing eigen\-values
of the Faddeev-Popov determinant within the first Gribov horizon, 
has not met with the widest consensus.
Indeed, the point has repeatedly been made\cite{Hirschfeld,Gov1}
that one ought rather to count the multiple intersections of the gauge 
slice with the selected gauge orbits with an alternating signature
determined by the oriented integration measure over the space of gauge orbits. 
Recently, that specific issue has been addressed again with the same conclusion
in contradistinction to Gribov's suggestion, within the context of
a solvable U(1) gauge invariant quantum mechanical model inspired
essentially by the dimensional reduction of ordinary Yang-Mills theories
to 0+1 dimensions\cite{Lee}.

Besides the Faddeev reduced phase space approach, there also exists the
BFV-BRST invariant extended phase space formulation of gauge invariant
systems, which in effect also imports the same issues of a gauge
fixing procedure and the ensuing possible Gribov problems\cite{Gov1}.
Even though these questions then arise in another disguise, nevertheless
they need to be addressed specifically within that context as well,
even in the simplest of examples\footnote{Note that when Gribov ambiguities
arise for a given gauge fixing procedure, be it in the Faddeev reduced phase 
space approach or the BFV-BRST invariant one, they need to be addressed
already at the classical level\cite{Gov1}.}\cite{Gov1}.

The issue of Gribov problems thus needs to be considered on a case by case
basis, not only for each physical system being studied, but also for each
gauge fixing procedure which may be contemplated for a given system.
Moreover, the Gribov problem issue is also very much dependent on the
choice of boundary conditions which are being implemented for a given
dynamical problem\cite{Gribov,Singer}. Hence, no procedure for resolving 
Gribov ambiguities in general, when they appear, can be formulated. 

However, such issues arise specifically because of the
apparent necessity of gauge fixing in gauge invariant systems.
If gauge fixing could be avoided altogether, no Gribov problems would
arise, and the issue would no longer need to be addressed. Indeed, the recent
proposal\cite{Klauder1} of the physical projector within Dirac's original 
approach to the quantisation of constrained systems\cite{Gov1} precisely
achieves\cite{Gov2} a correct integration over the space of gauge
orbits of a system, in which all orbits are effectively included once and
only once, without the necessity in no way
whatsoever of performing a gauge fixing procedure of any kind.
In particular when considering the time evolution operator for gauge invariant
states, the physical projector readily ensures that only physical states 
contribute as intermediate states to the physical propagator, a feature 
which is usually achieved only through gauge fixing to a reduced phase space
description or by extending the quantum dynamics to include
a ghost sector which compensates for those contributions from gauge
variant states.

The purpose of the present paper is to consider these different issues 
within the context of the simple U(1) gauge invariant solvable model 
of Ref.\cite{Lee}, by relying on a series of considerations and results
developed in Refs.\cite{Gov1,Victor,Klauder2,Gov3}. The definition of the
model, which is very much similar to the general class of systems studied
in Ref.\cite{Klauder2} already in terms of the physical projector,
is recalled in Sect.\ref{Sect2} together with its classical
Hamiltonian formulation. Sect.\ref{Sect3} then discusses its Dirac
quantisation, by constructing the configuration space wave functions
of the physical gauge invariant states. In Sect.\ref{Sect4}, a specific
admissible gauge fixing procedure leading to a reduced phase space
description of the model is developed, thereby enabling the explicit
evaluation of the configuration space representation of the gauge invariant
quantum evolution operator. Sect.\ref{Sect5} then considers the same
matrix elements within the BFV-BRST approach, and illustrates how
only admissible gauge fixing procedures---in the sense of the word
as defined above---lead to the correct result for the gauge invariant
evolution operator of physical states. All these results are then
directly contrasted against those obtained in Sect.\ref{Sect6}
through the physical projector approach simply defined within the context 
of the Dirac quantisation of the model, thereby readily leading again to 
the correct result for the physical propagator of the system to which only
physical states contribute as intermediate states. 
Finally, Sect.\ref{Sect7} presents the conclusions of the analysis.

\section{The Model and its Classical Hamiltonian Analysis}
\label{Sect2}

The dynamics of the U(1) gauge invariant model of Ref.\cite{Lee} is
defined by the Lagrangian,
\begin{equation}
L=\frac{1}{2}\left[(\dot{x}+g\xi y)^2+(\dot{y}-g\xi x)^2+
(\dot{z}-\xi)^2\right]-V\left(\sqrt{x^2+y^2}\right)\ \ \ ,
\label{eq:Lcar}
\end{equation}
or in terms of polar coordinates,
\begin{equation}
L=\frac{1}{2}\dot{r}^2+\frac{1}{2}r^2\left(\dot{\theta}-g\xi\right)^2+
\frac{1}{2}\left(\dot{z}-\xi\right)^2-V(r)\ \ \ ,
\label{eq:Lpolar}
\end{equation}
with of course,
\begin{equation}
x=r\cos\theta\ \ \ ,\ \ \ y=r\sin\theta\ \ \ .
\end{equation}
Here, $x(t)$, $y(t)$ and $z(t)$ are cartesian coordinates, $\xi(t)$ is
a gauge variable---essentially the time component of a U(1) gauge connection
after dimensional reduction to 0+1 dimensions---, and $g$ is a gauge 
coupling constant. Henceforth,
we also choose to work with the harmonic potential term,
\begin{equation}
V(r)=\frac{1}{2}\omega^2\,r^2\ \ \ .
\end{equation}

It should be clear that this model possesses a U(1) gauve invariance
whereby the $(x(t),y(t))$ coordinates are rotated by some arbitrary 
time dependent angle $\alpha(t)$ in the two dimensional plane which 
they define, while at the same time the variables $z(t)$ and $\xi(t)$ are 
shifted by a quantity proportional either to $\alpha(t)$ or $\dot{\alpha}(t)$.
Hence, this invariance of the system is best expressed in the polar
parametrisation, namely,
\begin{equation}
r'(t)=r(t)\ \ ,\ \ 
\theta'(t)=\theta+\alpha(t)\ \ ,\ \ 
z'(t)=z(t)+\frac{1}{g}\alpha(t)\ \ ,\ \ 
\xi'(t)=\xi(t)+\frac{1}{g}\dot{\alpha}(t)\ \ .
\end{equation}

This property of the system already raises the issue of the choice
of boundary conditions to be imposed on its degrees of freedom. Since
later on we are interested in computing the quantum evolution operator of
the model for gauge invariant states, boundary conditions need to be specified
at two distinct moments in time $t_i$ and $t_f$, with $t_f>t_i$. 
However, which of the
degrees of freedom $(x,y,z;\xi)$ need to have their values specified at these
values of $t$ is a matter of their physical status.

In fact, two distinct physical interpretations of the degrees of freedom 
$(x,y,z)$ are possible, among which two combinations are manifestly gauge 
invariant, namely,
\begin{equation}
r(t)\ \ \ ,\ \ \ \varphi(t)=\theta(t)-g z(t)\ \ \ ,
\end{equation}
while the third gauge {\em variant\/} combination may be chosen to be
for example,
\begin{equation}
\beta(t)=\theta(t)+gz(t)\ \ \ .
\end{equation}
In particular, note how the gauge invariant combinations $r$ and $\varphi$
show that the gauge orbits in the space $(x,y,z)$ are nothing else than
helicoidal curves of constant radius $r$, whose symmetry axis is parallel to
the $z$ axis, and whose slope w.r.t. the $(x,y)$ plane is set by the 
coupling $g$. Consequently, given any configuration $(r(t),\theta(t),z(t))$, 
a finite and unique gauge transformation of parameter function $\alpha(t)$ 
may always be found such that at all times $z'(t)=0$, namely
$\alpha(t)=-gz(t)$, so that also $\theta'(t)=\varphi(t)$.

Hence, one possible physical interpretation of the degrees of freedom of
the system is to consider that its actual configuration space is the entire
set of such helicoidal curves thus parametrised by the variables
$0\le r<+\infty$ and $0\le\varphi<2\pi$, and which are in one-to-one
correspondence with the points $(r,\theta,z=0)$ for all possible values 
of $0\le r<+\infty$ and $0\le\theta<2\pi$. This is the point of view 
taken in Ref.\cite{Lee} with regards to the nature of the configuration 
space of the system.
Associated to the specific choice of admissible gauge fixing effected through
the constraint $z(t)=0$, which leaves no room for further non trivial
gauge transformations, the appropriate choice of boundary conditions is thus,
\begin{equation}
r(t_i)=r_i\ \ ,\ \ \theta(t_i)=\theta_i\ \ ,\ \ z(t_i)=0\ \ ;\ \ 
r(t_f)=r_f\ \ ,\ \ \theta(t_f)=\theta_f\ \ ,\ \ z(t_f)=0\ \ ,
\label{eq:bc1}
\end{equation}
such that in particular $\varphi(t_{i,f})=\theta_{i,f}$. Note
that consistency with these boundary conditions requires the gauge
transformation parameter $\alpha(t)$ to vanish at the end points,
{\sl i.e.} $\alpha(t_{i,f}) = 0$.

However, this situation suggests another possibility, in which the physical
interpretation of the degrees of freedom is rather to consider that the
actual configuration space of the system is indeed the set of all coordinates
$(x,y,z)$ in a three dimensional euclidean space, with the two sets of
triplet values $\Big(r(t_{i,f}),\theta(t_{i,f}),z(t_{i,f})\Big)$ set to
specific boundary values $\Big(r_{i,f},\theta_{i,f},z_{i,f}\Big)$,
knowing that any physical quantity computed with this choice of boundary
conditions will depend only on the quantities  
$\Big(r_{i,f},\varphi_{i,f}=\theta_{i,f}-gz_{i,f}\Big)$.
In this second interpretation, one decouples so to say the gauge
transformations of the system for parameter functions $\alpha(t)$ which
vanish at the endpoints in time $t_{i,f}$ from those which do not
(a distinction which, within this interpretation, is consistent with 
the given boundary conditions), all in a continuous fashion.
The latter transformations may be used to transform the boundary conditions
such that $z(t)$ would vanish at the end points $t_{i,f}$. However, since we are
solely interested in the propagator for physical states, the latter quantity
may only depend on the gauge invariant combinations
$r_{i,f}$ and $\varphi_{i,f}$ of the boundary values of trajectories.
In other words, the physical propagator takes its values already only over
the actual configuration space of gauge orbits of the first interpretation
for the degrees of freedom of the model. Hence, in as far as the calculation
of the physical propagator is concerned, the action of those gauge
transformations for which $\alpha (t_{i,f}) \ne 0$ is already accounted
for through the dependency on the variables $\varphi_{i,f}$, rather than 
$\theta_{i,f}$ and $z_{i,f}$ separately. Thus given this second point
of view, gauge equivalence classes of physical trajectories are characterized 
by having fixed end points at $t=t_{i,f}$, while all their other points
$(x(t),y(t),z(t))$ associated to instants $t$ distinct from $t_{i,f}$
are gauge transformed into one another with arbitrary parameter functions
$\alpha(t)$ such that $\alpha(t_{i,f})=0$.

Except for Sect.\ref{Sect4} where the specific gauge fixing $z(t)=0$ is
used from the outset, we shall thus consider the following choice of boundary
conditions
\begin{equation}
r(t_i)=r_i\ \ ,\ \ \theta(t_i)=\theta_i\ \ ,\ \ z(t_i)=z_i\ \ ;\ \ 
r(t_f)=r_f\ \ ,\ \ \theta(t_f)=\theta_f\ \ ,\ \ z(t_f)=z_f\ \ ,
\label{eq:bc2}
\end{equation}
to which the second interpretation of the degrees of freedom $(x,y,z)$
is associated. Indeed, the structure of the Hamiltonian gauge invariance
of the model then becomes much similar to that of the parametrised
relativistic scalar particle\cite{Gov1}, 
so that results established in the latter case may
readily be borrowed for the calculation of the physical propagator in the
present system, and for a discussion of Gribov problems in the context of its
BFV-BRST invariant quantisation.

Let us now consider the Hamiltonian formulation of the system in its polar 
representation. Conjugate momenta are simply,
\begin{equation}
p_r=\dot{r}\ \ ,\ \ p_\theta=r^2\left[\dot{\theta}-g\xi\right]\ \ ,\ \ 
p_z=\dot{z}-\xi\ \ ,
\end{equation}
while there appears the usual primary constraint
\begin{equation}
p_\xi=0\ \ .
\end{equation}
The canonical Hamiltonian also reads,
\begin{equation}
H_0=\frac{1}{2}p^2_r+\frac{1}{2}\frac{1}{r^2}p^2_\theta+
\frac{1}{2}p^2_z+\frac{1}{2}\omega^2r^2+\xi\left[p_z+gp_\theta\right]\ \ \ .
\end{equation}
The consistent time evolution of the primary contraint $p_\xi=0$ then
requires only one more secondary constraint
\begin{equation}
\phi\equiv p_z+gp_\theta=0\ \ \ ,
\end{equation}
which together with the primary constraint $p_\xi=0$ forms a system of
first-class constraints, whose Poisson bracket is
\begin{equation}
\{p_\xi,p_z+gp_\theta\}=0\ \ \ .
\end{equation}

Equivalently\cite{Gov1}, this whole Hamiltonian formulation may simply
be specified through the associated first-order Lagrangian
\begin{equation}
L_1=\dot{r}p_r+\dot{\theta}p_\theta+\dot{z}p_z+\dot{\xi}p_\xi-
\frac{1}{2}p^2_r-\frac{1}{2}\frac{1}{r^2}p^2_\theta-
\frac{1}{2}p^2_z-\frac{1}{2}\omega^2r^2-\xi\left[p_z+gp_\theta\right]
-\lambda^1p_\xi-\lambda^2\left[p_z+gp_\theta\right]\ \ \ ,
\label{eq:L1}
\end{equation}
where $\lambda^{1,2}(t)$ are Lagrange multipliers for the two first-class
constraints. However, the following redefinitions of the Lagrange multipliers
\begin{equation}
\lambda^1=\dot{\xi}\ \ \ ,\ \ \
\tilde{\lambda}^2=\lambda^2+\xi\rightarrow \xi\ \ \ ,
\end{equation}
lead to the first-order Lagrangian
\begin{equation}
L_2=\dot{r}p_r+\dot{\theta}p_\theta+\dot{z}p_z-
\frac{1}{2}p^2_r-\frac{1}{2}\frac{1}{r^2}p^2_\theta-
\frac{1}{2}p^2_z-\frac{1}{2}\omega^2r^2-\xi\left[p_z+gp_\theta\right]\ \ \ .
\label{eq:L2}
\end{equation}

That such a reduction of the Hamiltonian formulation is consistent with
the gauge invariances generated by the first-class constraints $p_\xi$
and $\phi=p_z+gp_\theta$ has been demonstrated in Ref.\cite{Gov1}, thereby
leading to the so-called\cite{Gov1} ``fundamental Hamiltonian formulation"
of the present model. Indeed, the first-class constraint $p_\xi=0$ is a direct
consequence of the au\-xi\-lia\-ry character of the degree of freedom $\xi(t)$
which in fact turns out to be simply the Lagrange multiplier for the
second first-class constraint $\phi=0$, generator of the U(1) gauge symmetry
in the Hamiltonian formulation. Thus, $\xi(t)$ is actually not
a genuine dynamical degree of freedom of the system. By using the
fundamental Hamiltonian formulation, this fact is made explicit, while
using the first-order Lagrangian (\ref{eq:L1}) would introduce a new gauge
symmetry generated by $p_\xi$ whose sole purpose is to lead to arbitrary 
time dependent shifts in the variable $\xi(t)$, then artificially
considered as a genuine dynamical degree of freedom, while on the other hand
two more Lagrange multipliers $\lambda^{1,2}$ are then 
introduced\footnote{Were one to view the Lagrange multipliers $\lambda^{1,2}$
as genuine degrees of freedom\cite{Gov1}, further first-class constraints 
and their associated Lagrange multipliers would ensue, generating in turn
new Hamiltonian gauge symmetries similar to that associated to $\xi$ as 
just described. Thus, the dynamical status of $\xi$ is not different from 
that of $\lambda^{1,2}$, and the genuine dynamical content of the model is 
indeed solely represented by the dynamical evolution of the degrees of freedom 
$(r,\theta,z)$ and the constraint $\phi=p_z+gp_\theta$, namely the fundamental 
Hamiltonian formulation of the system, a fact which applies to all gauge 
invariant systems\cite{Gov1}.}.

Let us note at this point that a similar discussion applies to the
original Lagrangian formulation of the model in terms of its cartesian
coordinate parametrisation. In that case, the first-order Lagrangian
associated to its fundamental Hamiltonian formulation simply reads,
\begin{equation}
L_3=\dot{x}p_x+\dot{y}p_y+\dot{z}p_z-\frac{1}{2}p^2_x-\frac{1}{2}p^2_y-
\frac{1}{2}p^2_z-\frac{1}{2}\omega^2\left(x^2+y^2\right)-
\xi\left[p_z+g(xp_y-yp_x)\right]\ \ \ .
\label{eq:L3}
\end{equation}
However, because of the nature of the U(1) gauge invariance of the system,
the polar parametrisation is far more convenient than the cartesian
one, and most of our considerations are based on the former.

In its fundamental Hamiltonian description, the model thus possesses
three pairs of ca\-no\-ni\-cal\-ly conjugate phase space coordinates,
$(r,p_r)$, $(\theta,p_\theta)$ and $(z,p_z)$, subject to the first-class
constraint $\phi=p_z+gp_\theta=0$ of Lagrange multiplier $\xi$, and whose 
dynamics is generated by the first-class Hamiltonian
\begin{equation}
H=\frac{1}{2}p^2_r+\frac{1}{2}\frac{1}{r^2}p^2_\theta+
\frac{1}{2}p^2_z+\frac{1}{2}\omega^2r^2\ \ \ ,\ \ \ 
\{H,\phi\}=0\ \ \ .
\label{eq:H}
\end{equation}
In particular, the constraint $\phi$ is the generator of the U(1) gauge
symmetry on phase space through the symplectic structure defined by the
Poisson brackets. Infinitesimal transformations may be exponentiated to
finite ones, given by
\begin{equation}
r'=r\ \ ,\ \ p'_r=p_r\ \ ;\ \ 
\theta'=\theta+\alpha\ \ ,\ \ p'_\theta=p_\theta\ \ ;\ \ 
z'=z+\frac{1}{g}\alpha\ \ ,\ \ p'_z=p_z\ \ ;\ \ 
\xi'=\xi+\frac{1}{g}\dot{\alpha}\ \ ,
\label{eq:gauge1}
\end{equation}
where $\alpha(t)$ is an arbitrary time dependent function parametrising
Hamiltonian U(1) gauge transformations, subject to the
boundary conditions $\alpha(t_{i,f})=0$ when either of the choices of boundary
conditions (\ref{eq:bc1}) or (\ref{eq:bc2}) applies.

Since the gauge transformations of the Lagrange multiplier $\xi$ are independent
of the phase space degrees of freedom, the general notion of Teichm\"uller
space, namely the set of gauge orbits in the space of the Lagrange
multipliers associated to all first-class constraints\cite{Gov1}, applies
in the present instance. Given the boundary conditions $\alpha(t_{i,f})=0$,
it is clear that the following quantity,
\begin{equation}
\gamma=\int_{t_i}^{t_f}dt\,\xi(t)\ \ \ ,
\end{equation}
does define a gauge invariant quantity in the space of Lagrange multiplier
functions $\xi(t)$. Hence, $\gamma$ is a coordinate parametrising the
Teichm\"uller space of the system, which in the present case is
identified with the entire real line. Moreover, it should be quite clear
that any gauge fixing in the space of Lagrange multiplier functions
$\xi(t)$ induces a gauge fixing of the entire system itself in its
phase space formulation, since we must have $\alpha(t_{i,f})=0$. 
As a matter of fact, it may easily be shown\cite{Gov1} 
that any admissible gauge fixing in the space $\xi(t)$, thus associated
to a single covering of Teichm\"uller space in which all values of 
$\gamma$ are obtained once and only once when accounting for the
oriented integration measure on Teichm\"uller space, induces an admissible
gauge fixing of the entire system itself. Thus for example, a choice
of gauge fixing in the space $\xi(t)$ such that the following set of
functions is selected,
\begin{equation}
\xi(t;\gamma)=\frac{\gamma}{t_f-t_i}\ \ \ ,
\label{eq:gammasimple}
\end{equation}
where $\gamma$ is a free parameter taking all possible real values once and
only once, automatically induces an admissible gauge fixing of the
system itself. Hence, for the present model, gauge fixing may be considered
not only in terms of the dynamical degrees of freedom $(x,y,z)$, as shown
above through the choice $z(t)=0$ for example, but it may also be effected 
through gauge fixing in the space of Lagrange multiplier functions $\xi(t)$. 
Note also how the characterization of these different gauge fixing procedures 
and the possible ensuing Gribov problems is strongly dependent on the choice of
boundary conditions considered for the study of the time evolution of
system configurations.

In the polar parametrisation, the Hamiltonian equations of motion 
are\footnote{Incidentally, both for (\ref{eq:L2}) and (\ref{eq:L3}),
the Hamiltonian reduction whereby all conjugate phase space coordinates,
$p_r$, $p_\theta$ and $p_z$ in the first case, $p_x$, $p_y$ and $p_z$
in the second, are solved for through the Hamiltonian equations of motion,
leads back precisely to the original Lagrangian formulations
(\ref{eq:Lcar}) and (\ref{eq:Lpolar}) of the model.}
\begin{equation}
\dot{r}=p_r\ \ ,\ \ \dot{p}_r=\frac{p^2_\theta}{r^3}-\omega^2\,r\ \ ;\ \
\dot{\theta}=\frac{p_\theta}{r^2}+g\xi\ \ ,\ \ \dot{p}_\theta=0\ \ ;\ \ 
\dot{z}=p_z+\xi\ \ ,\ \ \dot{p}_z=0\ \ ,
\label{eq:HEM}
\end{equation}
subject further to the first-class constraint $\phi=p_z+gp_\theta=0$.
Note how due to the {\em global\/} symmetries of the model under decoupled
rotations in the $(x,y)$ plane and translations along the $z$ axis,
the angular momentum $p_\theta$ and the linear momentum $p_z$ are
each separately conserved quantities through time evolution,
$p_\theta(t)=L$ and $p_z(t)=p$. However,
the gauge invariance constraint $\phi=0$ defining physical configurations,
which generates time dependent translations in $z$ coupled to time
dependent rotations in $(x,y)$, brings into further relation these two 
conserved quantities, such that $p+gL=0$.

Given the choice of boundary conditions (\ref{eq:bc2}), the construction
of the general solution to these equations of motion proceeds as follows.
First, the values for the rotational energy\footnote{The total energy 
$E$ of the system is then given by $E=E_r+p^2/2$.} $E_r$ and for the angular
momentum $L$ must be determined such that\footnote{With our choice
of harmonic potential $V(r)=\omega^2r^2/2$, these conditions may be
solved explicitely, but the ensuing expressions are not very
informative, and are thus not given.},
\begin{equation}
t_f-t_i=\int_{r_i}^{r_f}\,du\,\frac{\pm 1}
{\sqrt{2\left[E_r-V(u)-\frac{L^2}{2u^2}\right]}}\ \ ,\ \ 
L=\frac{\varphi_f-\varphi_i}{g^2(t_f-t_i)+
\int_{t_i}^{t_f}\,\frac{dt}{r^2(t)}}\ \ \ ,
\end{equation}
where the solution for $r(t)$ is implicitely defined by the integral,
\begin{equation}
t-t_i=\int_{r_i}^{r(t)}\,du\,\frac{\pm 1}
{\sqrt{2\left[E_r-V(u)-\frac{L^2}{2u^2}\right]}}\ \ .
\end{equation}
In both this latter expression as well as in the previous one
particularised to $t=t_f$, the $\pm 1$ sign under the integral
stands for the sign of the time derivative $dr(t)/dt$.
Once the values for $E_r$ and $L$ thereby determined, the remaining phase space
variables are given by,
\begin{displaymath}
p_r(t)=\dot{r}(t)\ \ \ ,
\end{displaymath}
\begin{displaymath}
\theta(t)=\theta_i+L\int_{t_i}^t\,dt'\,\frac{1}{r^2(t')}+
g\int_{t_i}^t\,dt'\,\xi(t')\ \ ,\ \ p_\theta(t)=L\ \ \ ,
\end{displaymath}
\begin{equation}
z(t)=z_i-gL(t-t_i)+\int_{t_i}^t\,dt'\,\xi(t')\ \ ,\ \ 
p_z(t)=p=-gL\ \ \ .
\end{equation}

In view of the gauge transformation of the Lagrange multiplier in 
(\ref{eq:gauge1}), note how the function $\xi(t)$ appearing in these 
solutions parametrises the gauge freedom, {\em i.e.\/} the gauge redundancy,
of the general solutions to the equations of motion of the original
Euler-Lagrange equations associated to the Lagrangian (\ref{eq:Lpolar}).
In particular, the gauge invariant combination $\varphi(t)=\theta(t)-gz(t)$,
given by
\begin{equation}
\varphi(t)=\left[\theta_i-gz_i\right]+L
\int_{t_i}^t\,dt'\,\left(\frac{1}{r^2(t')}\,+\,g^2\right)\ \ \ ,
\end{equation}
is indeed independent of the Lagrange multiplier function $\xi(t)$.

However, for consistency of the construction of the
solution, the choice of Lagrange multiplier $\xi(t)$ must be such that
the associated Teichm\"uller parameter $\gamma$ takes the value,
\begin{equation}
\gamma=\int_{t_i}^{t_f}\,dt\,\xi(t)=(z_f-z_i)+gL(t_f-t_i)\ \ \ .
\label{eq:constraintbc}
\end{equation}
Hence, if a gauge fixing procedure is implemented such that the Teichm\"uller
parameter values $\gamma$ of the Lagrange multipliers $\xi(t)$ thereby 
effectively selected do not include this specific value required
by the choice of boundary conditions, the set of gauge orbits of the
system retained through gauge fixing does not include the specific one which
solves the equations of motion with this choice of boundary conditions.
Clearly, for an admissible gauge fixing, such a situation does never arise
since all possible values of $\gamma$ are then included once and only once.
This simple remark thus illustrates, at the classical level already,
how a non admissible gauge fixing procedure, thus suffering either 
a local or a global Gribov problem, 
excludes from the retained configurations of the system or
includes with too large a multiplicity 
certain subsets which physically are perfectly acceptable and should thus
remain accessible to the system throughout its dynamical evolution
with a single multiplicity for each of its gauge orbits.

The set of Teichm\"uller parameter values selected through the admissible gauge
fixing choice $z(t)=0$ will be discussed in detail in Sect.\ref{Sect4}.
Generally, one-parameter sets of functions $\xi(t)$ with their associated 
values for the Teichm\"uller parameter
$\gamma$ may be obtained through any gauge fixing procedure
leading to an equation of the form\cite{Teit,Gov4,Gov1}
\begin{equation}
\dot{\xi}=F(\xi)\ \ \ ,
\label{eq:diffxi}
\end{equation}
where $F(\xi)$ is some given function. Indeed, the solution 
$\xi(t;\xi_f)$ to this condition
is determined in terms of a single integration constant, which for later
convenience we choose to be the value taken by $\xi(t;\xi_f)$ at $t=t_f$, namely
$\xi_f=\xi(t_f;\xi_f)$. Correspondingly, as the value of $\xi_f$ runs
from $-\infty$ to $+\infty$, the Teichm\"uller
parameter $\gamma(\xi_f)=\int_{t_i}^{t_f}dt\xi(t;\xi_f)$ takes its
values in a certain domain ${\cal D}_\gamma[F]$ and with a certain covering
of that domain which both directly depend on the function $F(\xi)$ in 
(\ref{eq:diffxi}). Within this framework, an admissible gauge fixing
is related to a function $F(\xi)$ such that the domain ${\cal D}_\gamma[F]$
is the entire real line and with a covering such that all values for $\gamma$
are obtained once and only once.

Since the respective Teichm\"uller spaces are identical, let us consider some 
of the specific examples discussed in the case of the parametrised relativistic 
scalar particle in Refs.\cite{Gov1,Gov4}. The choice
\begin{equation}
F(\xi)=a\xi+b\ \ \ ,
\label{eq:F1}
\end{equation}
implies,
\begin{equation}
\xi(t;\xi_f)=\left(\xi_f+\frac{b}{a}\right)e^{a(t-t_f)}-\frac{b}{a}\ \ ,\ \ 
\gamma(\xi_f)=\frac{1}{a}\left(\xi_f+\frac{b}{a}\right)\left(1-e^{-a(t_f-t_i)}
\right)-\frac{b}{a}(t_f-t_i)\ \ ,
\end{equation}
which clearly thus defines an admissible choice of gauge fixing, since as
$\xi_f$ varies from $-\infty$ to $+\infty$ the
corresponding domain ${\cal D}_\gamma[F]$ is then indeed the entire real line
with each value of $\gamma$ obtained once and only once, irrespectively of the
values of the arbitrary parameters $a$ and $b$ defining the function
$F(\xi)$. In particular, the simplest choice of such an admissible gauge
fixing is obtained with $F(\xi)=0$\cite{Teit}, thereby leading to the set of
Lagrange multiplier functions $\xi(t)$ mentioned in (\ref{eq:gammasimple})
with $\gamma(\xi_f)=\xi_f(t_f-t_i)$.

A choice of a quadratic function for $F(\xi)$ however, is associated to a non
admissible gauge fixing. Indeed\cite{Gov1}, even though all real values
of the Teichm\"uller parameter are then obtained, they are obtained twice
while $\xi_f$ runs from $-\infty$ to $+\infty$, {\em and with opposite
orientations\/}. Hence effectively, the actual covering of Teichm\"uller space
which is implied by such a choice for $F(\xi)$ vanishes, establishing the
non admissibility of such a gauge fixing, which thus possesses
a global Gribov problem, no gauge orbit being effectively retained.
Nevertheless, as the coefficient of the quadratic term in $F(\xi)$ vanishes, 
the previous class of admissible gauge fixings is recovered\cite{Gov1}.

The only other example borrowed from Ref.\cite{Gov1} we shall mention here is
\begin{equation}
F(\xi)=a\xi^3\ \ \ ,\ \ \ a>0\ \ \ .
\label{eq:F3}
\end{equation}
Correspondingly, one finds
\begin{equation}
\xi(t;\xi_f)=\frac{\xi_f}{\sqrt{1+2a\xi_f^2(t-t_f)}}\ \ ,\ \ 
\gamma(\xi_f)=\frac{1}{a\xi_f}\left[\sqrt{1+2a\xi^2_f(t_f-t_i)}-1\right]\ \ \ .
\end{equation}
Consequently, for all non vanishing positive values of the parameter $a$, 
this choice leads to a non admissible gauge fixing of the system. Indeed, as
$\xi_f$ runs from $-\infty$ to $+\infty$, the associated domain in 
Teichm\"uller space reduces to the finite interval 
${\cal D}_\gamma[F]=[-\sqrt{2(t_f-t_i)/a},\sqrt{2(t_f-t_i)/a}]$ 
with a single covering. In other words, even though it does not suffer a local
Gribov problem, this gauge fixing is non admissible 
since it suffers a global one. Nevertheless, 
in the the limit where the parameter $a$ vanishes, the admissible
gauge fixing implied by $F(\xi)=0$ is indeed recovered.

Incidentally, note that even already at the classical level, these simple
examples illustrate the general fact that the gauge invariant physics 
described by gauge invariant systems {\em is not independent\/} of the
gauge fixing procedure to which they are subjected\cite{Gov1,Gov4}, 
contrary to what seems to be
generally believed to be true. Gauge invariance of physical quantities
is not all there is to gauge invariant systems; it is also imperative
that the actual space of gauge orbits of the system be properly accounted for
by any description based on some gauge fixing procedure. This can only be
achieved through an admissible gauge fixing, even though any gauge fixing
procedure, {\em including those suffering local or global Gribov 
problems\/}, always leads to gauge invariant results for physical observables.

\section{Dirac Quantisation}
\label{Sect3}

Dirac's quantisation of the model simply consists in the canonical
quantisation of the previous Hamiltonian formulation of the system,
with the constraint of gauge invariance $\phi=0$ imposed at the
operatorial level in order to define physical, {\sl i.e.\/} gauge invariant
states. For convenience, we choose to work in the polar parametrisation
$(r,\theta,z)$ of the degrees of freedom of the system, which thus
requires the construction of representations of the associated Heisenberg
algebra in these curvilinear coordinates parametrising
the three dimensional euclidean space defined by the cartesian 
coordinates $(x,y,z)$. Moreover, given our intent of computing 
the propagator of physical states, we shall consider from the outset
the configuration space representation of the canonical
commutation relations in polar coordinates.

For this purpose, we rely on the recent classification\cite{Gov3} 
of representations of the Heisenberg algebra in arbitrary coordinate systems 
set up on arbitrary manifolds. In the present instance, since the euclidean
space $(x,y,z)$ is simply connected, only the trivial representation
of the Heisenberg algebra exists, while the differential operator representation
of the conjugate momenta ope\-ra\-tors $\hat{p}_r$, $\hat{p}_\theta$ and 
$\hat{p}_z$ is determined by the metric structure of this configuration space, 
expressed in polar coordinates by
\begin{equation}
ds^2=dr^2+r^2d\theta^2+dz^2\ \ \ .
\end{equation}
Consequently\cite{Gov3}, the configuration space wave function inner product 
is defined by the integration measure
\begin{equation}
\int_0^{+\infty}dr\ r\ \int_0^{2\pi}d\theta\ \int_{-\infty}^{+\infty}dz\ \ \ ,
\end{equation}
while the conjugate momenta operators are represented by the
differential operators,
\begin{equation}
\hat{p}_r:\ \ -\frac{i\hbar}{\sqrt{r}}\partial_r\sqrt{r}\ \ \ ;\ \ \
\hat{p}_\theta:\ \ -i\hbar\partial_\theta\ \ \ ;\ \ \
\hat{p}_z:\ \ -i\hbar\partial_z\ \ \ ,
\end{equation}
the coordinate operators $\hat{r}$, $\hat{\theta}$ and $\hat{z}$ 
being of course represented on configuration space wave functions through 
simple multiplication by the associated eigenvalues $r$, $\theta$ and $z$.

This does not yet specify the choice of quantum Hamiltonian for the
system, in cor\-res\-pondence with the classical first-class Hamiltonian
$H$ in (\ref{eq:H}). However, the canonical choice corresponding to the
usual scalar Laplacian differential operator, is defined by\cite{Gov3}
\begin{equation}
\hat{H}=\frac{1}{2}\frac{1}{\sqrt{\hat{r}}}\hat{p}_r\,
\hat{r}\,\hat{p}_r\,\frac{1}{\sqrt{\hat{r}}}\,+\,
\frac{1}{2}\frac{1}{\hat{r}^2}\hat{p}^2_\theta\,+\,\frac{1}{2}\hat{p}^2_z\,+\,
\frac{1}{2}\omega^2\hat{r}^2\ \ \ ,
\end{equation}
which, in terms of the above representation of the conjugate momenta operators,
also reads,
\begin{equation}
\hat{H}:\ \ \ -\frac{\hbar^2}{2}\left[
\partial^2_r+\frac{1}{r}\partial_r+\frac{1}{r^2}\partial^2_\theta+
\partial^2_z\right]\,+\,\frac{1}{2}\omega^2r^2\ \ \ .
\end{equation}

Finally, the gauge generator operator $\hat{\phi}=\hat{p}_z+g\hat{p}_\theta$ 
is thus also represented by
\begin{equation}
\hat{\phi}:\ \ \ -i\hbar\left(\partial_z+g\partial_\theta\right)\ \ \ .
\end{equation}
In particular, by definition, the configuration space wave functions of 
gauge invariant states are annihilated by this latter differential operator.

Let us first consider the diagonalisation of the first-class Hamiltonian 
$\hat{H}$, whose set of eigenstates thus defines a basis for the full space of
quantum states of the system, of which a specific linear subspace is
that of the gauge invariant physical states annihilated by the gauge
generator $\hat{\phi}$. In the same manner as at the classical level, 
since $\hat{H}$, $\hat{p}_\theta$ and $\hat{p}_z$ are all commuting operators, 
a common diagonalisation of all these three operators may be found, thereby 
also diagonalising the gauge generator $\hat{\phi}$. Thus, a basis for the 
physical states of the system is defined by the specific subset of this 
diagonalising basis of states whose $\hat{\phi}$ eigenvalue vanishes identically.

With these considerations in mind, the resolution of the Schr\"odinger equation
\begin{equation}
\Big\{-\frac{\hbar^2}{2}\left[
\partial^2_r+\frac{1}{r}\partial_r+\frac{1}{r^2}\partial^2_\theta+
\partial^2_z\right]\,+\,\frac{1}{2}\omega^2r^2\Big\}\psi(r,\theta,z)=
E\psi(r,\theta,z)\ \ \ ,
\end{equation}
is straightforward enough. The eigenvalue spectrum is given by
\begin{equation}
E_{m,\ell,p}=\frac{1}{2}p^2+\hbar\omega\Big[2m+|\ell|+1\Big]\ \ \ ,
\end{equation}
with $-\infty<p<+\infty$, $m=0,1,2,\dots$ and $\ell=0,\pm 1,\pm 2,\dots$,
while the corresponding eigenstate configuration space wave functions
are
\begin{displaymath}
\psi_{m,\ell,p}(r,\theta,z)=<r,\theta,z|m,\ell,p>=
\end{displaymath}
\begin{equation}
=(-1)^m\,
\left(\frac{\omega}{\pi\hbar}\right)^{1/2}\,
\left(\frac{1}{2\pi\hbar}\right)^{1/2}\,
\left(\frac{m!}{(m+|\ell|)!}\right)^{1/2}\,
e^{i\ell\theta}\,e^{ipz/\hbar}\,u^{|\ell|}\,e^{-u^2/2}\,L^{|\ell|}_m(u^2)\ \ \ ,
\label{eq:wave}
\end{equation}
where $L^{|\ell|}_m(x)$ are the usual Laguerre polynomials, the
variable $u$ is defined by $u=r\sqrt{\omega/\hbar}$ and the choice of
overall phase $(-1)^m$ is made for later convenience. The normalisation
of these states is such that
\begin{equation}
<m,\ell,p|m',\ell',p'>=\delta_{mm'}\,\delta_{\ell\ell'}\,\delta(p-p')\ \ \ .
\end{equation}

In particular, a basis for the physical states of the model is provided
by these wave functions with the further restriction that their $\hat{\phi}$
eigenvalue vanishes, namely
\begin{equation}
p=-\hbar g\ell\ \ \ .
\end{equation}
Note that the dependency of the associated wave functions on the gauge
variant variables $\theta$ and $z$ does combine into a single
phase dependency on the specific gauge invariant
combination $\varphi=\theta-gz$, namely
with the following recombination of phase factors
\begin{equation}
e^{i\ell\theta}\,e^{ipz/\hbar}\ \rightarrow\
e^{i\ell(\theta-gz)}=e^{i\ell\varphi}\ \ \ ,
\end{equation}
as was indeed expected.

This explicit resolution of the quantised model
may of course also be achieved through algebraic methods\cite{Lee,Klauder2}.
For that purpose, it is more appropriate to consider first the
cartesian parametrisation of its degrees of freedom $(x,y,z)$, and the
definition of the usual creation and annihilation operators (only the latter
are recalled here),
\begin{equation}
a_1=\sqrt{\frac{\omega}{2\hbar}}\left[\hat{x}+\frac{i}{\omega}\hat{p}_x\right]
\ \ \ ,\ \ \ 
a_2=\sqrt{\frac{\omega}{2\hbar}}\left[\hat{y}+\frac{i}{\omega}\hat{p}_y\right]
\ \ \ .
\end{equation}
However, rotational invariance of the model in the $(x,y)$ plane calls
for the introduction of an helicity-like basis of creation and
annihilation operators, defined by\cite{Klauder2},
\begin{equation}
a_\pm=\frac{1}{\sqrt{2}}\left[a_1\mp ia_2\right]\ \ \ ,\ \ \ 
a_\pm^\dagger=\frac{1}{\sqrt{2}}\left[a_1^\dagger\pm 
ia_2^\dagger\right]\ \ \ .
\end{equation}

In term of these quantities, the first-class Hamiltonian and the gauge generator
read,
\begin{equation}
\hat{H}=\frac{1}{2}\hat{p}^2_z+\hbar\omega\left[a_+^\dagger a_+\,+\,
a_-^\dagger a_-\right]\ \ \ ,\ \ \ 
\hat{\phi}=\hat{p}_z+\hbar g\left[a_+^\dagger a_+\,-\,
a_-^\dagger a_-\right]\ \ \ .
\end{equation}
Hence, the orthonormalised helicity Fock state basis, extended here to 
$\hat{p}_z$ momentum eigenstates and defined by
\begin{equation}
|n_\pm,p>=\frac{1}{\sqrt{n_+!\ n_-!}}\,\left(a_+^\dagger\right)^{n_+}\,
\left(a_-^\dagger\right)^{n_-}\,|0,p>\ \ \ ,
\end{equation}
and with the Fock vacuum $|0,p>$ being such that
\begin{equation}
a_\pm\,|0,p>=0\ \ \ ,\ \ \ \hat{p}_z|0,p>=p\,|0,p>\ \ \ ,\ \ \
<0,p|0,p'>=\delta(p-p')\ \ \ ,
\end{equation}
provides a direct diagonalisation both of the Hamiltonian $\hat{H}$ and of the
gauge generator $\hat{\phi}$. The energy spectrum is thus
\begin{equation}
\hat{H}|n_\pm,p>=E_{n_\pm,p}\,|n_\pm,p>\ \ \ ,\ \ \ 
E_{n_\pm,p}=\frac{1}{2}p^2+\hbar\omega\left(n_++n_-+1\right)\ \ \ ,
\end{equation}
while the physical state condition $\hat{\phi}=0$ leads to the restriction
\begin{equation}
p=-\hbar g\left(n_+-n_-\right)\ \ \ .
\end{equation}

To establish complete identity with the previous results, one only needs
now to determine the configuration space wave function representation
of the states $|n_\pm,p>$. This is a straightforward exercise using the
differential operator representations of the operators $\hat{p}_r$ and 
$\hat{p}_\theta$ introduced previously. One then finds\cite{Lee} precisely 
the same wave functions as those given in (\ref{eq:wave}), with the following 
correspondence
\begin{equation}
<r,\theta,z|n_\pm,p>=\psi_{m,\ell,p}(r,\theta,z)\ \ \ ,\ \ \ 
m={\rm min}(n_+,n_-)\ \ \ ,\ \ \ \ell=n_+-n_-\ \ \ .
\end{equation}

Since the first-class Hamiltonian commutes with the gauge generator,
any physical state retains this quality under time evolution generated
by the evolution operator of the system given by the time-ordered
expression
\begin{equation}
\hat{U}(t_f,t_i)=Te^{-\frac{i}{\hbar}\int_{t_i}^{t_f}\,
dt'\,\left[\hat{H}+\xi(t')\hat{\phi}\right]}\ \ \ ,
\label{eq:evolDirac}
\end{equation}
where $\xi(t)$ stands for an arbitrary choice of Lagrange multiplier.
Nevertheless, this fact does not allow a direct evaluation of the
configuration space matrix elements of the evolution operator
to which only physical states would contribute as intermediate states,
since configuration space eigenstates $|r,\theta,z>$ do not define
gauge invariant states. Hence, it appears that in order to evaluate
the configuration space propagator of the system restricted to gauge
invariant states only, one needs first to implement some gauge fixing
procedure by which the contributions from all gauge {\em variant\/} 
variables are removed, while retaining only those of the physical
states. The calculation of this physical state propagator through
gauge fixing is the purpose of the next two sections, before
showing in Sect.\ref{Sect6} how the same goal may readily be reached 
by using the physical projector\cite{Klauder1} simply
within Dirac's quantisation of the system which does not require
any gauge fixing whatsoever.

\section{Reduced Phase Space Quantisation}
\label{Sect4}

In this section, we consider Faddeev's reduced phase space 
formulation\cite{Gov1} of the model given the gauge fixing condition 
$z(t)=0$, to which the choice of boundary conditions (\ref{eq:bc1}) 
thus applies. As was discussed previously, we know that this choice of 
gauge fixing is admissible\footnote{Other reduced phase space gauge fixings,
such as $z(t)-\lambda x(t)=0$ with $\lambda$ a fixed postive parameter, 
are discussed in Ref.\cite{Lee}.}. Hence, the canonical quantisation of the 
corresponding Hamiltonian formulation should allow for the calculation 
of the physical propagator of the system, to which all gauge orbits 
of the model contribute once and only once.

Together with the first-class generator $\phi=p_z+gp_\theta$
of the U(1) gauge symmetry, the gauge fixing condition
\begin{equation}
\Omega=z=0\ \ \ ,
\end{equation}
does define a set of second-class constraints, whose Faddeev-Popov
determinant does not vanish,
\begin{equation}
\{\phi,\Omega\}=-1\ \ \ .
\end{equation}
In particular, the requirement that the gauge fixing condition $\Omega=0$
be maintained under time evolution generated by the total Hamiltonian
$H_T=H+\xi\phi$, namely $\dot{\Omega}=0$, implies the following specification
of the Lagrange multiplier,
\begin{equation}
\xi=-p_z=gp_\theta\ \ \ .
\end{equation}
Hence, given the fact that $p_\theta(t)$ keeps a constant value $L$ for
solutions to the equations of motion, the associated Teichm\"uller
parameter
\begin{equation}
\gamma=\int_{t_i}^{t_f}dt\,\xi(t)=gL(t_f-t_i)\ \ \ ,
\end{equation}
does indeed obey the constraint (\ref{eq:constraintbc}) set by
the boundary conditions (\ref{eq:bc1}). Of course, the fact that this
constraint is met is a consequence of the admissibility of the gauge fixing 
condition $z(t)=0$.

The reduction of the system, and in particular the decoupling of the
$(z,p_z)$ sector of phase space degrees of freedom through the relations
\begin{equation}
z=0\ \ \ ,\ \ \ p_z=-gp_\theta\ \ \ ,
\end{equation}
necessitates the calculation of the Dirac brackets associated to the 
second-class constraints $(\phi,\Omega)$. One easily finds that the Dirac 
brackets among the remaining phase space conjugate pairs $(r,p_r)$ and 
$(\theta,p_\theta)$ are identical to their original Poisson brackets.
In addition, the reduced Hamiltonian appropriate to this reduced
phase space formulation of the model is given by
\begin{equation}
H_{\rm red}=\frac{1}{2}p^2_r\,+\,\frac{1}{2}\left[\frac{1}{r^2}+g^2\right]
p^2_\theta\,+\,\frac{1}{2}\omega^2r^2\ \ \ .
\end{equation}
In particular, the corresponding equations of motion,
\begin{equation}
\dot{r}=p_r\ \ \ ,\ \ \ \dot{p}_r=\frac{1}{r^3}p^2_\theta\,-\,\omega^2r
\ \ \ ;\ \ \ 
\dot{\theta}=\left(\frac{1}{r^2}+g^2\right)p_\theta\ \ \ ,\ \ \
\dot{p}_\theta=0\ \ \ ,
\end{equation}
coincide with those obtained from the original equations of motion
(\ref{eq:HEM}) in which the relations $z=0$, $p_z=-gp_\theta$
and $\xi=-p_z=gp_\theta$ are being substituted.

The canonical quantisation of this formulation of the model in
polar coordinates is straightforward enough, since it coincides
with that of the $(r,\theta)$ sector in Dirac's quantisation.
Hence, we may immediately transcribe the corresponding configuration
space representations of quantum operators. In particular,
a basis of the quantum space of states, which are now necessarily
all gauge invariant ones, is provided by those states which diagonalise
both the angular momentum operator $\hat{p}_\theta=-i\hbar\partial_\theta$
and the Schr\"odinger operator associated to the reduced Hamiltonian
above, namely,
\begin{equation}
\Big\{-\frac{\hbar^2}{2}\left[
\partial^2_r+\frac{1}{r}\partial_r+
\left(\frac{1}{r^2}+g^2\right)\partial^2_\theta\right]\,+\,
\frac{1}{2}\omega^2r^2\Big\}\psi(r,\theta)=
E\psi(r,\theta)\ \ \ .
\end{equation}
The explicit resolution of this equation shows that the eigenvalue spectrum 
is given by
\begin{equation}
E_{m,\ell}=\frac{1}{2}\hbar^2g^2\ell^2+
\hbar\omega\Big[2m+|\ell|+1\Big]\ \ \ ,
\end{equation}
with $m=0,1,2,\dots$ and $\ell=0,\pm 1,\pm 2,\dots$,
while the corresponding eigenstate configuration space wave functions
are
\begin{equation}
\psi_{m,\ell}(r,\theta)=<r,\theta|m,\ell>=(-1)^m\,
\left(\frac{\omega}{\pi\hbar}\right)^{1/2}\,
\left(\frac{m!}{(m+|\ell|)!}\right)^{1/2}\,
e^{i\ell\theta}\,u^{|\ell|}\,e^{-u^2/2}\,L^{|\ell|}_m(u^2)\ \ \ ,
\end{equation}
with the same notations as in (\ref{eq:wave}) and a normalisation such that
\begin{equation}
<m,\ell|m',\ell'>=\delta_{mm'}\,\delta_{\ell\ell'}\ \ \ .
\end{equation}
Except for the normalisation factor $1/(2\pi\hbar)^{1/2}$ stemming 
from the $(z,p_z)$ sector which has been
reduced in the present approach, this energy spectrum as well as these
wave functions coincide with those established for physical states
in Dirac's quantisation (recall that we have here $z=0$ so that
$\varphi=\theta-gz=\theta$).

The calculation of the physical propagator is now immediate. Since it
is defined by the matrix elements
\begin{equation}
P_{\rm red}(i\rightarrow f)=<r_f,\theta_f|
e^{-\frac{i}{\hbar}(t_f-t_i)\hat{H}_{\rm red}}|r_i,\theta_i>\ \ \ ,
\end{equation}
we explicitely have
\begin{equation}
P_{\rm red}(i\rightarrow f)=\sum_{m=0}^{+\infty}\sum_{\ell=-\infty}^{+\infty}
<r_f,\theta_f|m,\ell>\,e^{-\frac{i}{\hbar}(t_f-t_i)E_{m,\ell}}\,
<m,\ell|r_i,\theta_i>\ \ \ .
\end{equation}
Given the wave functions established above, the summation over 
the integer $m$ is possible in terms of Bessel functions of the first
kind, leading finally to the following expression for the configuration
space propagator of physical states\footnote{Had the factor
${\rm exp}\left(-i\hbar\Delta t g^2\ell^2/2\right)$ been absent 
from this expression,
the summation over $\ell$ would have been possible as well\cite{Gov3}, leading
of course to the usual propagator for the two dimensional spherically
symmetric harmonic oscillator of angular frequency $\omega$ and unit 
mass $m=1$.},
\begin{equation}
P_{\rm red}(i\rightarrow f)=\frac{\omega}{2i\pi\hbar\sin\omega\Delta t}\,
e^{\frac{i}{2}\frac{\cos\omega\Delta t}{\sin\omega\Delta t}(u^2_f+u^2_i)}\,
\sum_{\ell=-\infty}^{+\infty}\,e^{-i\frac{\pi}{2}|\ell|}\,
e^{-\frac{i}{2}\hbar\Delta t g^2\ell^2}\,
e^{i\ell(\theta_f-\theta_i)}\,
J_{|\ell|}\left(\frac{u_f u_i}{\sin\omega\Delta t}\right)\ \ \ ,
\label{eq:propred}
\end{equation}
where $\Delta t=t_f-t_i$. This latter result will thus serve as the
point of comparison for the physical propagator obtained through the other
two quantisation approaches considered in this paper, namely the
so-called BFV-BRST invariant formulation of gauge invariant systems,
and the physical projector construction of Ref.\cite{Klauder1}.

\section{BFV-BRST Quantisation}
\label{Sect5}

As opposed to the reduced phase space approach, the BFV-BRST one\cite{Gov1}
extends the set of dynamical degrees of freedom in the following manner.
First, the Lagrange multiplier $\xi(t)$ is promoted to being a dynamical
variable by introducing its conjugate momentum $p_\xi(t)$, thereby leading
now to two first-class constraints
\begin{equation}
G_{a=1}=p_\xi=0\ \ \ ,\ \ \ G_{a=2}=\phi=p_z+gp_\theta=0\ \ \ .
\end{equation}
Note that we still have
\begin{equation}
\{G_a,G_b\}=0\ \ \ ,\ \ \ \{H,G_a\}=0\ \ \ ,\ \ \ a,b=1,2\ \ \ .
\end{equation}

Next, in order to compensate for these additional degrees of freedom,
further dynamical variables of opposite Grassmann parity are introduced,
the BFV ghosts $\eta^a(t)$ and ${\cal P}_a(t)$, $(a=1,2)$, each such pair
being canonically conjugate variables with graded Poisson brackets
\begin{equation}
\{\eta^a,{\cal P}_b\}=-\delta^a_b\ \ \ ,\ \ \ a,b=1,2\ \ \ .
\end{equation}
In addition, the BFV ghosts have the following properties under complex
conjugation,
\begin{equation}
\left(\eta^a\right)^*=\eta^a\ \ \ ,\ \ \ 
\left({\cal P}_a\right)^*=-{\cal P}_a\ \ \ ,\ \ \ a=1,2\ \ \ .
\end{equation}

Within this framework, the original Hamiltonian gauge invariance generated
by the first-class constraint $\phi=G_2$ is now traded for a global
BRST symmetry generated by the Grassmann odd BRST charge $Q_B$, which for 
the present model is given by
\begin{equation}
Q_B=\eta^a\,G_a=\eta^1\,p_\xi\,+\,\eta^2\left[p_z+gp_\theta\right]\ \ \ ,
\end{equation}
and is characterised by the nilpotency property $\{Q_B,Q_B\}=0$
as well as being real under complex conjugation, $\left(Q_B\right)^*=Q_B$.

Similarly, the original first-class Hamiltonian $H$ may also be extended
to a BRST invariant one, $H_B$, which in the present case is identical 
to $H$, $H_B=H$. However, in the same way that the time evolution of the
system in Dirac's construction is generated by the Hamiltonian $H$ to
which an arbitrary linear combination of the first-class constraints is
added, here as well time evolution in this extended phase space
description of the system is generated by the most general possible
BRST invariant Hamiltonian based on $H_B$, given by
\begin{equation}
H_{\rm eff}=H_B-\{\Psi,Q_B\}\ \ \ .
\end{equation}
Here, $\Psi$ is {\em a priori\/} a totally arbitrary function on the
extended phase space, of Grassmann odd parity and odd under complex
conjugation, while the nilpotency property of the BRST charge ensures that
\begin{equation}
\{H_{\rm eff},Q_B\}=0\ \ \ .
\end{equation}

Given any choice for the function $\Psi$, the equations of motion of the
system within this extended framework are easily established. In order
to construct gauge invariant solutions however, it is imperative to
impose BRST invariant boundary conditions which extend the choice considered
previously in (\ref{eq:bc2}). A general discussion\cite{Gov1} shows 
that the following conditions always meet 
this requirement of BRST invariance,
\begin{equation}
p_\xi(t_i)=0\ \ ,\ \ {\cal P}_1(t_i)=0\ \ ,\ \ \eta^2(t_i)=0\ \ ;\ \ 
p_\xi(t_f)=0\ \ ,\ \ {\cal P}_1(t_f)=0\ \ ,\ \ \eta^2(t_f)=0\ \ ,
\label{eq:bcBRST}
\end{equation}
a fact which in the present case is explicitely confirmed by considering
the BRST transformations of the extended phase space variables,
\begin{displaymath}
\delta_B r=\{r,Q_B\}=0\ ,\
\delta_B p_r=0\ ;\ 
\delta_B\theta=g\eta^2\ ,\ 
\delta_B p_\theta=0\ ;\
\delta_B z=\eta^2\ ,\ \delta_B p_z=0\ ;\
\end{displaymath}
\begin{equation}
\delta_B\xi=\eta^1\ ,\ 
\delta_B p_\xi=0\ ;\
\delta_B\eta^1=0\ ,\ 
\delta_B{\cal P}_1=-p_\xi\ ;\
\delta_B\eta^2=0\ ,\
\delta_B{\cal P}_2=-\left[p_z+gp_\theta\right]\ .
\end{equation}

In the remainder of this section, we shall consider the specific 
choice\cite{Gov1}
\begin{equation}
\Psi=F(\xi){\cal P}_1+\xi{\cal P}_2\ \ \ ,
\label{eq:Psi}
\end{equation}
where $F(\xi)$ is an arbitrary function. The corresponding effective
Hamiltonian is then
\begin{equation}
H_{\rm eff}=\frac{1}{2}p^2_r+\frac{1}{2r^2}p^2_\theta+\frac{1}{2}p^2_z
+\frac{1}{2}\omega^2r^2+\xi\Big[p_z+gp_\theta\Big]+
F(\xi)p_\xi-F'(\xi){\cal P}_1\eta^1-{\cal P}_2\eta^1\ \ \ ,
\end{equation}
from which it is possible to derive the equations of motion for all
the extended phase space variables $(r,p_r)$, $(\theta,p_\theta)$,
$(z,p_z)$, $(\xi,p_\xi)$ and $(\eta^a,{\cal P}_a)$. Among these equations,
the one for the Lagrange multiplier is simply
\begin{equation}
\dot{\xi}=F(\xi)\ \ \ .
\end{equation}
In other words, within the BFV-BRST framework, the function $\Psi$
provides the ``gauge fixing fermion function", which for the choice
in (\ref{eq:Psi}) implies a gauge fixing of the system in its
space of Lagrange multiplier functions $\xi(t)$ precisely of the type
discussed in (\ref{eq:diffxi}), for which admissible and non admissible
examples were described.

Rather than considering now the construction of the general solution
to these equations of motion at the classical level, let us turn immediately
to the BRST quantisation of the system. Through the redefinitions
\begin{equation}
c^a=\hat{\eta}^a\ \ \ ,\ \ \ b_a=\frac{i}{\hbar}\hat{{\cal P}}_a
\ \ \ ,\ \ \ a=1,2\ \ \ ,
\end{equation}
the operator algebra in the ghost sector is specified by the set of
anticommutation relations
\begin{equation}
\{c^a,b_b\}=\delta^a_b\ \ ,\ \ \left(c^a\right)^2=0\ \ ,\ \ 
\left(b_a\right)^2=0\ \ ,\ \ 
{c^a}^\dagger=c^a\ \ ,\ \ b^\dagger_a=b_a\ \ ,
\ \ a,b=1,2\ \ ,
\end{equation}
thus defining the usual so-called $(b,c)$ ghost system whose representation
theory is straightforward and well known\cite{Gov1}.
Further, the definitions of the BRST and gauge fixing fermion operators,
$\hat{Q}_B$ and $\hat{\Psi}$, read as those at the classical level above, while
the BRST invariant effective quantum Hamiltonian operator $\hat{H}_{\rm eff}$
is now given by,
\begin{equation}
\hat{H}_{\rm eff}=\hat{H}+\frac{i}{\hbar}\{\hat{\Psi},\hat{Q}_B\}\ \ \ ,
\end{equation}
where the choice of normal ordering in the $p_r^2$ contribution
to the first-class Hamiltonian operator $\hat{H}$ is that of Sect.\ref{Sect3}.
Explicitely, one finds,
\begin{equation}
\hat{H}_{\rm eff}=\frac{1}{2}\hat{p}^2_r-\frac{\hbar^2}{8\hat{r}^2}
+\frac{1}{2\hat{r}^2}\hat{p}^2_\theta
+\frac{1}{2}\hat{p}^2_z+\frac{1}{2}\omega^2\hat{r}^2
+\hat{\xi}\Big[\hat{p}_z+g\hat{p}_\theta\Big]+
F(\hat{\xi})\hat{p}_\xi-i\hbar F'(\hat{\xi})c^1b_1-i\hbar c^1b_2\ \ .
\end{equation}

We are now in a position to compute the physical propagator of the system,
namely the matrix elements of the BRST invariant evolution operator
\begin{equation}
\hat{U}_{\rm BRST}(t_f,t_i)=
e^{-\frac{i}{\hbar}\Delta t \hat{H}_{\rm eff}}\ \ \ ,
\end{equation}
for the BRST invariant external states of the quantised system which
correspond to the choice of BRST invariant boundary conditions 
in (\ref{eq:bc2}) and (\ref{eq:bcBRST}). However, since the extended sectors 
of variables $(\xi,p_\xi)$ and $(\eta^a,{\cal P}_a)$ as well as the structure 
of the Hamiltonian gauge algebras of the present model and of the parametrised
relativistic scalar particle are identical, only the outline of the
calculation will be discussed here, whose complete details are thoroughly
presented in Ref.\cite{Gov1} for the latter system. In particular,
for reasons presented in that work, the calculation of the matrix elements
we are interested in, is best performed using the path integral representation
over the BFV extended phase space. The construction of a discretized
but exact expression for that path integral proceeds in the usual fashion,
by considering the $(N-1)$-times insertion of the spectral decomposition of the
identity operator on the space of quantum states between the $N$ factors
appearing in the following rewriting of the evolution operator,
\begin{equation}
\hat{U}_{\rm BRST}(t_f,t_i)=
\left[e^{-\frac{i}{\hbar}\epsilon \hat{H}_{\rm eff}}\right]^N\ \ \ ,\ \ \
\epsilon=\frac{\Delta t}{N}=\frac{t_f-t_i}{N}\ \ \ .
\end{equation}
Using wave function representations for the position and momentum
eigenstates associated to all extended phase space operators 
$(\hat{r},\hat{p}_r)$, $(\hat{\theta},\hat{p}_\theta)$, 
$(\hat{z},\hat{p}_z)$, $(\hat{\xi},\hat{p}_\xi)$ and 
$(\hat{\eta}^a,\hat{{\cal P}}_a)$,
this procedure leads to a discretized representation of the extended 
phase space path integral associated to the relevant matrix element of the
BRST invariant evolution operator in configuration space (further details
are found in Ref.\cite{Gov1}). In particular, the ghost sector is
represented in terms of Grassmann odd variables, whose integration
then combines with that over the Lagrange multiplier sector $(\xi,p_\xi)$
and may be completed exactly.

Having performed the integrations over the extended sector of degrees
of freedom $(\xi,p_\xi)$ and $(\eta^a,{\cal P}_a)$, the path integral
representation of the BRST invariant physical propagator 
reads\footnote{Compared to Ref.\cite{Gov1}, the only difference is that
the $(r,\theta)$ sector is that of curvilinear coordinates in an euclidean
space, for which the considerations developed in Ref.\cite{Gov3} must be
applied. This is the reason for the factor $1/\sqrt{r_f\,r_i}$ in front
of this expression, while the phase $i$ stems from the inner product
in the $(c^a,b_a)$ ghost sectors\cite{Gov1}.},
\begin{displaymath}
P^{[F]}_{\rm BRST}\left(i\rightarrow f\right)=\frac{i}{\sqrt{r_f\,r_i}}\,
\lim_{N\rightarrow\infty}\,
\int_0^{+\infty}\prod_{\alpha=1}^{N-1}dr_\alpha\,
\int_{-\infty}^{+\infty}\prod_{\alpha=0}^{N-1}\frac{dp_{r\alpha}}{2\pi\hbar}\,
\int_0^{2\pi}\prod_{\alpha=1}^{N-1}d\theta_\alpha\,
\prod_{\alpha=0}^{N-1}
\left[\frac{1}{2\pi}\sum_{\ell_\alpha=-\infty}^{+\infty}\right]\,\times
\end{displaymath}
\begin{displaymath}
\times\int_{-\infty}^{+\infty}\prod_{\alpha=1}^{N-1}dz_\alpha\,
\int_{-\infty}^{+\infty}\prod_{\alpha=0}^{N-1}\frac{dp_{z\alpha}}{2\pi\hbar}\,
\frac{1}{2\pi\hbar}\int_{-\infty}^{+\infty}\prod_{\alpha=0}^{N-1}d\xi_\alpha\,
\delta\Big(\xi_{\alpha-1}-\xi_\alpha+\epsilon F(\xi_\alpha)\Big)\,
\frac{d\gamma(\xi_\alpha)}{d\xi_{N-1}}\,\times
\end{displaymath}
\begin{equation}
\times {\rm exp}\Bigg\{\frac{i}{\hbar}\sum_{\alpha=0}^{N-1}\left[
(r_{\alpha+1}-r_\alpha)p_{r\alpha}+
\hbar(\theta_{\alpha+1}-\theta_\alpha)l_\alpha+
(z_{\alpha+1}-z_\alpha)p_{z\alpha}-\epsilon h_\alpha
-\epsilon\xi_\alpha\left(p_{z\alpha}+g\ell_\alpha\right)\right]\Bigg\}\ \ \ ,
\end{equation}
with
\begin{equation}
h_\alpha=\frac{1}{2}p^2_{r\alpha}-\frac{\hbar^2}{8r^2_\alpha}+
\frac{\hbar^2\ell^2_\alpha}{2r^2_\alpha}+\frac{1}{2}p^2_{z\alpha}+
\frac{1}{2}\omega^2r^2_\alpha\ \ \ ,\ \ \ 
\gamma(\xi_\alpha)=\epsilon\sum_{\alpha=0}^{N-1}\xi_\alpha\ \ \ ,
\end{equation}
and of course the boundary values
\begin{equation}
r_{\alpha=0}=r_i\ \ ,\ \ r_{\alpha=N}=r_f\ \ ;\ \ 
\theta_{\alpha=0}=\theta_i\ \ ,\ \ \theta_{\alpha=N}=\theta_f\ \ ;\ \ 
z_{\alpha=0}=z_i\ \ ,\ \ z_{\alpha=N}=z_f\ \ .
\end{equation}

Hence, in the limit $N\rightarrow\infty$, the extended sector of variables
$(\xi,p_\xi)$ and $(\eta^a,{\cal P}_a)$ does again lead to the gauge fixing
of the system implied by the differential equation (\ref{eq:diffxi})
and the function $F(\xi)$. In that limit, as a consequence of the BRST
invariance of the quantity being computed, the integration over that sector
does indeed lead to an integral over Teichm\"uller space parametrised
by the parameter $\gamma$ and with the measure $d\gamma/(2\pi\hbar)$,
but over a domain ${\cal D}_\gamma[F]$ determined from the differential
equation $d\xi/dt=F(\xi)$ and the boundary value $\xi_f=\xi(t_f)$
precisely in the manner discussed in Sect.\ref{Sect2}. Thus, even though
the quantity of interested is BRST and gauge invariant, and thus
defined over Teichm\"uller space rather than the space of Lagrange multiplier
functions $\xi(t)$, nevertheless this gauge invariant quantity
{\em is not independent of the gauge fixing procedure\/}\cite{Gov1,Gov4}. 
Indeed, as the examples of Sect.\ref{Sect2} have illustrated, 
the domain ${\cal D}_\gamma[F]$ obtained through the classes of gauge 
fixings implied by the
choice (\ref{eq:diffxi}) is dependent both on the type of function $F(\xi)$
and on the parameters defining that function. Once again, gauge invariance
of physical observables is not all there is to gauge invariant systems.

Further integrations lead to the expression
\begin{displaymath}
P^{[F]}_{\rm BRST}\left(i\rightarrow f\right)=
\frac{1}{2\pi}\sum_{\ell=-\infty}^{+\infty}\,e^{i\ell(\theta_f-\theta_i)}\,
\int_{-\infty}^{+\infty}\frac{dp}{2\pi\hbar}\,
e^{ip(z_f-z_i)/\hbar}\,e^{-ip^2\Delta t/(2\hbar)}\,\times
\end{displaymath}
\begin{displaymath}
\times\,\int_{{\cal D}_\gamma[F]}\frac{d\gamma}{2\pi\hbar}\,
e^{-\frac{i}{\hbar}\gamma(p+g\ell)}\,
\frac{i}{\sqrt{r_f\,r_i}}\,\lim_{N\rightarrow\infty}\,
\left(\frac{1}{2i\pi\hbar\epsilon}\right)^{N/2}\,
\int_0^{+\infty}\prod_{\alpha=1}^{N-1}dr_\alpha\,\times
\end{displaymath}
\begin{equation}
\times\,{\rm exp}\Bigg\{\sum_{\alpha=0}^{N-1}\left[
\frac{i}{2\hbar\epsilon}(r_{\alpha+1}-r_\alpha)^2-
\frac{1}{2}i\hbar\epsilon(\ell^2-\frac{1}{4})\frac{1}{r^2_\alpha}-
\frac{i\epsilon}{2\hbar}\omega^2r^2_\alpha\right]\Bigg\}\ \ \ .
\end{equation}

However, the integration over the radial variable $r$ is precisely that
which appears in the path integral calculation of the propagator
of the two dimensional spherically symmetric harmonic oscillator.
Using the techniques developed in Ref.\cite{Peak}, one then finally obtains
\begin{displaymath}
P^{[F]}_{\rm BRST}\left(i\rightarrow f\right)=
i\,\sum_{\ell=-\infty}^{+\infty}\,e^{i\ell(\theta_f-\theta_i)}\,
\int_{-\infty}^{+\infty}\frac{dp}{2\pi\hbar}\,
e^{ip(z_f-z_i)/\hbar}\,e^{-ip^2\Delta t/(2\hbar)}\,
\int_{{\cal D}_\gamma[F]}\frac{d\gamma}{2\pi\hbar}\,
e^{-\frac{i}{\hbar}\gamma(p+g\ell)}\,\times
\end{displaymath}
\begin{equation}
\times\,\frac{\omega}{2i\pi\hbar\sin\omega\Delta t}\,
e^{-i\frac{\pi}{2}|\ell|}\,
e^{\frac{i\omega}{2\hbar}\frac{\cos\omega\Delta t}{\sin\omega\Delta t}
(r^2_f+r^2_i)}\,
J_{|\ell|}\left(\frac{\omega r_f\,r_i}{\hbar\sin\omega\Delta t}\right)\ \ \ .
\end{equation}

In order to reduce further this latter result and compare it to the
exact expression (\ref{eq:propred}) obtained for the admissible
gauge fixing $z(t)=0$, it should be clear that a choice for the
function $F(\xi)$ must be made such that the corresponding gauge fixing
is also admissible, thereby leading to the domain ${\cal D}_\gamma[F]$ 
being the entire real line. For any non admissible choice for
$F(\xi)$, the associated domain ${\cal D}_\gamma[F]$ would not be
the entire Techm\"uller space of the system, and the final result
obtained for the nevertheless gauge invariant quantity
$P^{[F]}_{\rm BRST}\left(i\rightarrow f\right)$ would not
represent the correct expression (\ref{eq:propred}) 
for the physical propagator of
gauge invariant states of the model\footnote{For example, with the choice
$F(\xi)=a\xi^3$, the final result for the BRST invariant
propagator would depend on the parameter $a$ defining the gauge fixing
condition, and only in the limit $a\rightarrow 0$ would the admissible
result be recovered. Nevertheless, by construction, 
$P^{[F]}_{\rm BRST}(i\rightarrow f)$ is gauge invariant whatever the value 
for $a$, since the BFV-BRST path integral is gauge invariant independently 
of the choice for $F(\xi)$.}.

Thus assuming now that the function $F(\xi)$ defines an {\em admissible\/}
gauge fixing of the system\footnote{Such as for example
$F(\xi)=a\xi+b$.}---with the meaning defined in the Introduction---,
it is clear that finally, the exact evaluation of the exact path integral
representation of the BRST invariant propagator leads to the expression
\begin{displaymath}
P^{[{\rm admissible}\ F]}_{\rm BRST}\left(i\rightarrow f\right)=
\frac{i}{2\pi\hbar}\,
\frac{\omega}{2i\pi\hbar\sin\omega\Delta t}\,
e^{\frac{i\omega}{2\hbar}\frac{\cos\omega\Delta t}{\sin\omega\Delta t}
(r^2_f+r^2_i)}\,\times
\end{displaymath}
\begin{equation}
\times\,\sum_{\ell=-\infty}^{+\infty}\,e^{-i\frac{\pi}{2}|\ell|}\,
e^{i\ell(\varphi_f-\varphi_i)}\,
e^{-\frac{i}{2}\hbar\Delta t g^2\ell^2}\,
J_{|\ell|}\left(\frac{\omega r_f\,r_i}{\hbar\sin\omega\Delta t}\right)\ \ \ ,
\label{eq:propBRST}
\end{equation}
in which we have used the definitions $\varphi_{i,f}=\theta_{i,f}-gz_{i,f}$
of the gauge invariant combinations of the boundary conditions,
which indeed appear in this expression as initially expected for
a gauge invariant quantity.

Quite clearly, except for the overall normalisation factor
$i/(2\pi\hbar)$ stemming from the quantum dynamics of the
extended sector of variables $(\xi,p_\xi)$ and $(\eta^a,{\cal P}_a)$
in the BFV-BRST invariant framework, the final result in
(\ref{eq:propBRST}) {\em valid only for an admissible gauge fixing\/}
coincides {\em exactly\/} with the exact result in (\ref{eq:propred})
obtained within the reduced phase space approach based on the
admissible gauge fixing condition $z(t)=0$ for the same physical
propagator of the quantised model. In particular, this conclusion
provides an explicit demonstration of the well known fact that the
extended sector of Lagrange multiplier and ghost degrees of freedom
of opposite Grassmann pa\-ri\-ties, precisely cancels out the contributions
of the gauge variant states to the matrix elements of physical observables,
while only those contributions of physical states are retained.

\clearpage

\section{The Physical Projector}
\label{Sect6}

The previous two sections have demonstrated how through a rather
involved process of gauge fixing requiring a careful analysis of the possible
Gribov problems which may thereby ensue, the configuration space propagator
of the physical states of the quantised system may be obtained, but with
an expression which is physically correct {\em only for an admissible
choice of gauge fixing\/} free of any local or global Gribov ambiguity.
Both approaches rely first on Dirac's Hamiltonian formulation of
constrained systems, which is then either reduced or extended before
the quantum dynamics of the system is considered.

The purpose of this section is to illustrate that all these complications
may be avoided altogether, without the necessity of any gauge fixing
whatsoever, by working immediately within Dirac's quantisation of
constrained systems and exploiting the construction of the physical
projector onto gauge invariant states introduced in Ref.\cite{Klauder1}.
In particular, we shall reconsider once again the calculation of the
physical propagator for the choice of boundary conditions (\ref{eq:bc2}),
to explicitely establish that the correct result (\ref{eq:propred})
is indeed readily derived using such an approach. This analysis only
requires the results of Dirac's quantisation derived in Sect.\ref{Sect3}.

In that section, it was shown that the spectrum of the U(1) gauge
symmetry generator $\hat{\phi}$ is given by
\begin{equation}
\hat{\phi}:\ \ p+\hbar g\ell\ \ \ ,\ \ \ -\infty<p<+\infty\ \ ,\ \ 
\ell=0,\pm 1,\pm 2,\dots\ \ \ .
\end{equation}
This spectrum being continuous---due to the non compact component of the
gauge symmetry group related to the translations in the variable $z$ which
it induces---, the actual definition of the projection operator onto
physical states annihilated by the operator $\hat{\phi}$ requires\cite{Klauder1}
to consider a finite eigenvalue interval $[-\delta,\delta]$, with a positive
quantity $\delta$ taken as small as is wished. The projection operator
onto $\hat{\phi}$ eigenstates whose eigenvalue lies within that interval
is then given by\cite{Klauder1}
\begin{equation}
\proj_{\ \delta}=\int_{-\infty}^{+\infty}d\gamma\,
\frac{\sin(\delta\gamma/\hbar)}{\pi\gamma}\,e^{\frac{i}{\hbar}\gamma\hat{\phi}}
\ \ \ .
\end{equation}

Considering then the quantum evolution operator (\ref{eq:evolDirac}) in
Dirac's quantisation, its projection onto contributions from physical
states only is defined by
\begin{equation}
\hat{U}_{\rm phys}(t_f,t_i)=
\lim_{\delta\rightarrow 0}\,\frac{\pi\hbar}{\delta}\,\hat{U}(t_f,t_i)\,
\proj_{\ \delta}=
\lim_{\delta\rightarrow 0}\,\frac{\pi\hbar}{\delta}\,
e^{-\frac{i}{\hbar}\Delta t \hat{H}}\,\proj_{\ \delta}\ \ \ ,
\end{equation}
where $\hat{H}$ is the first-class quantum Hamiltonian of the system.
Note that since $\hat{H}$ and $\hat{\phi}$ commute, one may also write
\begin{equation}
e^{-\frac{i}{\hbar}\Delta t \hat{H}}\,\proj_{\ \delta}=
\proj_{\ \delta}\,e^{-\frac{i}{\hbar}\Delta t\,\proj_{\ \delta} 
\hat{H}\proj_{\ \delta}}\,
\proj_{\ \delta}\ \ \ ,
\end{equation}
in order to emphasize the fact that indeed only physical states contribute to 
the physical evolution operator $\hat{U}_{\rm phys}(t_f,t_i)$, both as 
intermediate and as external states, in the limit $\delta\rightarrow 0$.

Hence, the configuration space physical propagator is simply given by
the following matrix element\footnote{The coherent state matrix elements
of the same operator have been considered in Ref.\cite{Klauder1}.},
\begin{equation}
P_{\rm proj}\left(i\rightarrow f\right)=
<r_f,\theta_f,z_f|\hat{U}_{\rm phys}(t_i,t_f)|r_i,\theta_i,z_i>\ \ \ ,
\label{eq:DefPproj}
\end{equation}
whose explicit evaluation only requires the wave functions of a complete
basis of all quantum states of the system---including the non 
physical ones---constructed in (\ref{eq:wave}). Since this basis
diagonalises both the Hamiltonian $\hat{H}$ and the generator $\hat{\phi}$,
the calculation is rather straightforward and follows much the same
lines as the one in Sect.\ref{Sect4} for the summation over the
integer $m$ associated to the $(r,\theta)$ sector of degrees of freedom.
The additional contributions from the $z$ sector are easily included,
since the projector $\proj_{\ \delta}$ implies the constraint $p+\hbar g\ell=0$
through a $\delta(p+\hbar g\ell)$ function in the limit $\delta\rightarrow 0$.

To explicitely illustrate how the contributions of gauge variant states
as intermediate states are indeed projected out from the physical propagator, 
let us consider the expression for (\ref{eq:DefPproj}) when the complete set of 
eigenstate configuration space wave functions (\ref{eq:wave}) is substituted for,
\begin{displaymath}
P_{\rm proj}\left(i\rightarrow f\right)=
\sum_{m=0}^{+\infty}\sum_{\ell=-\infty}^{+\infty}\int_{-\infty}^{+\infty}dp\
\psi_{m,\ell,p}(r_f,\theta_f,z_f)\,\times
\end{displaymath}
\begin{equation}
\times\, 
e^{-\frac{i}{\hbar}\Delta t\left[\frac{1}{2}p^2+\hbar\omega(2m+|\ell|+1)\right]}
\,\int_{-\infty}^{+\infty}d\gamma\,e^{\frac{i}{\hbar}\gamma(p+\hbar g\ell)}
\,\psi^*_{m,\ell,p}(r_i,\theta_i,z_i)
\ \ \ .
\label{eq:propinter}
\end{equation}
In this form, it is clear that the integration over the Teichm\"uller parameter
$\gamma$ enforces the gauge invariance constraint through the factor
$2\pi\hbar\delta(p+\hbar g\ell)$. From this contribution, the numerical
factor $2\pi\hbar$ cancels precisely the factor $1/(2\pi\hbar)$ which stems
from the normalisation of the wave functions (\ref{eq:wave}) in the $(z,p_z)$
sector of degrees of freedom, while the $\delta(p+\hbar g\ell)$ function
implies that only those states for which $p=-\hbar g\ell$, namely physical
states, are retained in the summation over the integers $m$ and $\ell$
and the integration over the conserved momentum $p$.
Hence, the physical projector, leading to an integration over Teichm\"uller
space, does indeed project out the contributions of the gauge variant
states of the quantised system; only physical states contribute as intermediate
states to the physical propagator, all and every single one of them contributing
with the same multiplicity of unity.

The remainder of the calculation is then identical to that which leads
to the result (\ref{eq:propred}) established in the reduced phase space
approach. Indeed, the summation over the integers $m$ and $\ell$ is that
already performed in that case, with an identical normalisation of the
wave functions in the $(r,\theta)$ sector of degrees of freedom.
Hence, a direct enough and exact calculation readily provides the
following final expression for the physical propagator within the physical
projector approach,
\begin{displaymath}
P_{\rm proj}\left(i\rightarrow f\right)=
\frac{\omega}{2i\pi\hbar\sin\omega\Delta t}\,
e^{\frac{i\omega}{2\hbar}\frac{\cos\omega\Delta t}{\sin\omega\Delta t}
(r^2_f+r^2_i)}\,\times
\end{displaymath}
\begin{equation}
\times\,\sum_{\ell=-\infty}^{+\infty}\,e^{-i\frac{\pi}{2}|\ell|}\,
e^{i\ell(\varphi_f-\varphi_i)}\,
e^{-\frac{i}{2}\hbar\Delta t g^2\ell^2}\,
J_{|\ell|}\left(\frac{\omega r_f\,r_i}{\hbar\sin\omega\Delta t}\right)\ \ \ ,
\label{eq:propproj}
\end{equation}
a result which coincides {\em exactly\/} with those
in (\ref{eq:propred}) and (\ref{eq:propBRST}) established
on the basis of some gauge fixing procedure whose admissibility 
must be ascertained with great care, and by using methods going beyond
the simpler framework of Dirac's quantisation of constrained systems.

As it should, the physical projector has thus achieved the required projecting 
out of the contributions of the gauge variant states to the physical propagator,
independently of any gauge fixing procedure and thereby avoiding any
potential Gribov problem. With hindsight, the result (\ref{eq:propproj})
could have been obtained straighforwardly already in Sect.\ref{Sect3}
within Dirac's quantisation of the system\cite{Victor}, by restricting
by hand so to say, the summation in (\ref{eq:propinter})
over the intermediate states contributing
to the evolution operator (\ref{eq:evolDirac}) to the subspace of the
physical states only, namely
\begin{equation}
\sum_{m=0}^{+\infty}\sum_{\ell=-\infty}^{+\infty}\,
\psi_{m,\ell,p}(r_f,\theta_f,z_f)\,
e^{-\frac{i}{\hbar}\Delta t
\left[\frac{1}{2}\hbar^2 g^2\ell^2+\hbar\omega(2m+|\ell|+1)\right]}
\,\psi^*_{m,\ell,p}(r_i,\theta_i,z_i)
\ \ \ ,
\end{equation}
a relation in which the energy spectrum of gauge invariant states is also
accounted for. Except for an overall normalisation factor of $1/(2\pi\hbar)$ 
stemming from the $(z,p_z)$ sector of degrees of freedom, quite obviously the 
same result as those established above for the physical propagator is obtained.
Nevertheless, for systems not as simple as the present one, such a direct
restriction to physical state contributions only is not expected to be
feasable ``by hand" in such a straightforward way within Dirac's 
quantisation, while the physical projector which finds its rightful setting
within that very same framework, is in general the appropriate tool for that 
purpose which is then achieved directly albeit often implicitely.

\section{Conclusions}
\label{Sect7}

Using the simple but yet rich enough solvable U(1) gauge invariant quantum
mechanical model of Ref.\cite{Lee}, this work has demonstrated the clear
advantages of using the physical projector\cite{Klauder1} in the quantisation
of gauge invariant systems. Indeed, by its very definition, the physical 
projector avoids the apparent necessity of some gauge fixing procedure,
which most often is at the origin of local and global Gribov problems
rendering the associated gauge invariant description of a given system
physically in- or over-complete. Moreover, all methods of gauge fixing
require an approach going beyond the original Dirac Hamiltonian framework
for constrained systems, within which however, the physical projector finds 
its rightful setting.

As the present paper has clearly illustrated, the
physical projector naturally avoids all the complications inherent
to any gauge fixing procedure, both through the automatic lack of 
any Gribov problem and by the absence of any further developments beyond 
those required in any case by Dirac's approach. In addition, 
the physical projector induces
implicitely the correct integration of all gauge orbits of a gauge
invariant system\cite{Gov2}, by effectively accounting for the contributions
to the dynamics of the system---be it classical or quantum---of each one
of all the gauge orbits once and only once, the very fact which characterizes
precisely what should define an admissible gauge fixing procedure when one is
being implemented.

In particular, by considering the explicit calculation of the physical 
propagator, it was clearly demonstrated how the physical projector 
automatically enforces the fact that only gauge invariant physical states
contribute to that observable as intermediate states, the contributions
of the gauge variant states being simply projected out. Within the other
approaches to the quantisation of gauge invariant systems, which all
require some gauge fixing procedure and are thus potentially plagued by
Gribov problems, the cancellation of the gauge variant
contributions is achieved either by having only gauge invariant configurations
to survive the phase space reduction, or by having an extended sector
of ghost degrees of freedom to compensate for the contributions of gauge
variant configurations. It is only for an admissible gauge 
fixing that this cancellation of gauge variant contributions is correctly 
achieved, a feature which is directly though implicitely realised within 
the physical projector approach given the very character of the latter operator.

Quite obviously, the diverse advantages of the physical projector
in a gauge fixing free quantisation of gauge invariant systems are
there to be explored much further for systems whose dynamics is not 
as simple as that of the model used in this paper, beginning for example 
with Yang-Mills theory in 1+1 and 2+1 dimensions, or topological
quantum field theories\cite{Gov5}. And beyond such applications, 
it may hoped that this approach to the quantisation of gauge invariant 
systems will provide new insights\cite{Klauder3} 
into the mathematical and physical riches of the
actual gauge theories of the fundamental interactions.

\section{Acknowledgments}

This work has been partially supported by the grant
CONACyT 3979P-E9608 under the terms of an agreement
between the CONACyT (Mexico) and the FNRS (Belgium).


\clearpage

\newpage

\begin{thebibliography}{99}

\bibitem{Gribov} V.N. Gribov, {\sl Nucl. Phys.\/} {\bf B139} (1978) 1.

\bibitem{Singer} I.M. Singer, {\sl Comm. Math. Phys.\/} {\bf 60} (1978) 7.

\bibitem{Gov1} For a detailed discussion and references to the original
literature, see for example,\\
J. Govaerts, {\em Hamiltonian Quantisation and Constrained
Dynamics\/}, (Leuven University Press, Leuven, 1991).

\bibitem{Hirschfeld} K. Fujikawa, {\sl Prog. Theor. Phys.\/} 
{\bf 61} (1979) 627;\\
P. Hirschfeld, {\sl Nucl. Phys.\/} {\bf B157} (1979) 37;\\
M.B. Halpern and J. Koplik, {\sl Nucl. Phys.\/} {\bf B132} (1978) 239.

\bibitem{Lee} R. Friedberg, T.D. Lee, Y. Pang and H.C. Ren, {\sl Ann.
Phys.\/} {\bf 246} (1996) 381.

\bibitem{Klauder1} J.R. Klauder, {\sl Ann. Phys.\/} {\bf 254} (1997) 419;\\
J.R. Klauder, {\sl Nucl. Phys.\/} {\bf B547} (1999) 397.

\bibitem{Gov2} J. Govaerts, {\sl J. Phys.\/} {\bf A30} (1997) 603.

\bibitem{Victor} V.M. Villanueva, {\sl Cuantizaci\'on de sistemas singulares
en variedades Riemannianas\/},\\
Ph.D.~Thesis
(Instituto de F\'{\i}sica, Universidad de Guanajuato, May 1999, unpublished).

\bibitem{Klauder2}
J. Govaerts and J.R. Klauder, {\sl Ann. Phys.\/} {\bf 274} (1999) 251.

\bibitem{Gov3} J. Govaerts and V.M. Villanueva, {\em Topology Classes of
Flat U(1) Bundles and Diffeomorphic Covariant Representations of the
Heisenberg Algebra\/}, preprint {\tt quant-ph/9908014} (August 1999).

\bibitem{Teit} C. Teitelboim, {\sl Phys. Rev.\/} {\bf D25} (1982) 3159.

\bibitem{Gov4} J. Govaerts, {\sl Int. J. Mod. Phys.} {\bf A4} (1989) 173;\\
J. Govaerts, {\sl Int. J. Mod. Phys.} {\bf A4} (1989) 4487;\\
J. Govaerts and W. Troost, {\sl Class. Quantum Grav.} {\bf 8} (1991) 1723.

\bibitem{Peak} D. Peak and A. Inomata, {\sl J. Math. Phys.\/} {\bf 10}
(1969) 1422.

\bibitem{Gov5} J. Govaerts and B. Deschepper, {\sl The Physical Projector 
and Topological Quantum Field Theories: U(1) Chern-Simons Theory in 2+1 
Dimensions\/}, in preparation;\\
B. Deschepper, Diploma Thesis (Catholic University of
Louvain, June 1999).

\bibitem{Klauder3} S.V. Shabanov and J.R. Klauder, {\sl Phys. Lett.\/}
{\bf B456} (1999) 38.

\end{thebibliography}
\end{document}